\documentclass{iaus}
\usepackage{graphicx}

\title[Direct Imaging Planet Searches and Transiting Planets] 
{The Impact of Transiting Planet Science on the Next Generation
of Direct-Imaging Planet Searches}

\author[Joseph C. Carson]   
{Joseph C. Carson
 }

\affiliation{Max-Planck-Institut f$\ddot{u}$r Astronomie  \\ K$\ddot{o}$nigstuhl 17,
 69115 Heidelberg, Germany \\ email: {\tt jcarson@mpia.de} }

\pubyear{2008}
\volume{253}  
\jname{Transiting Planets}
\begin{document}

\maketitle

\begin{abstract}
Within the next five years, a number of direct-imaging planet search
instruments, like the {\it VLT SPHERE} instrument, will be coming 
online.  To successfully carry out their programs, these instruments will rely heavily on a-priori information on planet composition, atmosphere, and 
evolution.  Transiting planet surveys, while covering a different semi-major axis regime, have the potential to provide critical foundations for these next-generation surveys.  For example, improved information on planetary evolutionary tracks may significantly impact the insights that can be drawn from 
direct-imaging statistical data.  Other high-impact results from transiting planet science include information on mass-to-radius relationships as well as 
atmospheric absorption bands.  The marriage of transiting planet and direct-imaging results may eventually give us the first complete picture of planet 
migration, multiplicity, and general evolution. 

\end{abstract}

\firstsection 
\section{Examples of Upcoming Direct-Imaging Planet-Search Systems}

{\underline{{\bf Ground-Based}}}

{\it -  HiCIAO}: An adaptive optics coronagraphic simultaneous-differential-imager for 
the Subaru 8.2 meter telescope (Tamura et al. 2006);  currently in commissioning phase.

{\it -  Project 1640}:  An adaptive optics coronagraphic integral field spectrograph for the 
Palomar 5 meter Telescope (Hinkley et al. 2008);  currently in commissioning phase.

{\it -  SPHERE}:  An adaptive optics coronagraphic simultaneous-differential-imager and 
integral field spectrograph for one of the VLT 8.2 meter telescopes (Beuzit et al. 2008);  commissioning expected 2011. 

{\it -  GPI}:  An adaptive optics coronagraphic integral field spectrograph for the Gemini 
South 8 meter telescope (Macintosh et al. 2006);  commissioning expected 2011. 
   
{\underline{{\bf Space-Based}}}

{\it Terrestrial Planet Finder Coronagraph\footnote[1]{http://planetquest.jpl.nasa.gov/TPF-C}?}

{\it Terrestrial Planet Finder Interferometer\footnote[2]{http://planetquest.jpl.nasa.gov/TPF-I}/Darwin\footnote[3]{http://www.esa.int/science/darwin}?}

\vspace{4 mm}
All of these systems will rely on results from transiting planet science to accurately interpret their observations.

\section{How Transiting Planet Radius Estimates Affect Direct-Imaging Results}

     Direct-imaging surveys, that observe the thermal emission from planets, are unable to make a direct 
measurement of planet radius.   In order to convert an observed magnitude into a planet temperature they must 
therefore usually rely on a-priori knowledge of typical planet radii (for a given planet mass).  Results from 
transiting planet surveys (see compilation in Bakos et al. 2004 for instance), however, suggest that typical planet 
radii could be larger than researchers have previously presumed.  More data from transiting planet surveys are 
required to better establish this mass-radius relationship.  The conclusions from such data will be important for 
accurately determining the sensitivities of direct imaging programs.  The plot below shows how estimated direct-imaging sensitivities (minimum detectable mass as a function of separation from the parent star) may be affected by changes in planet radii assumptions             

\begin{figure}[b]
\begin{center}
\includegraphics[width=3.4in]{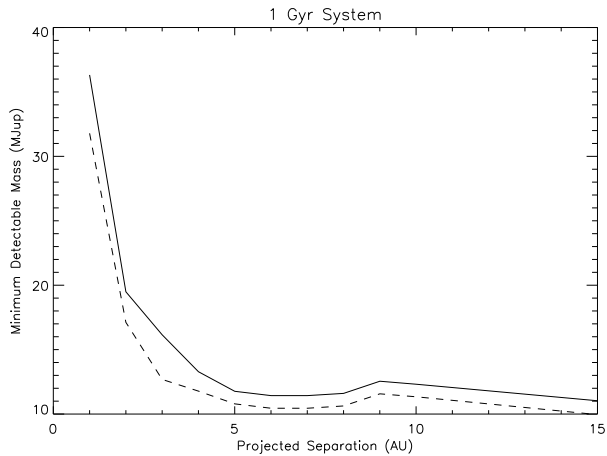}
\caption{     This plot shows how estimated (near-IR) direct-imaging high contrast 
sensitivities may be affected by errors in a-priori assumptions of typical planet radii. 
 The solid curve  assumes a conventional mass-radius relationship (e.g. a 1 M$_{Jup}$ 
planet has a 1 R$_{Jup}$ radius).  The dashed curve  corresponds to the case where actual 
typical planet radii are 25\% larger than classical models assume.  For the displayed 
1 Gyr system, the expected minimum detectable mass at 1 AU changes from $\sim$36 
M$_{Jup}$ to $\sim$32 M$_{Jup}$.  This disparity becomes smaller for younger systems and larger 
for older systems.  
     The plots assume a high-contrast sensitivity curve adapted from example
     {\it VLT 
SPHERE} simulations for a target system 10pc away.  Planet masses are derived 
from temperature and age using Baraffe et al. (2004) models.  For a given mass and 
age, all planets are assumed to have identical atmospheric optical depths.}  
\label{fig1}
\end{center}
\end{figure}

     Figure 2 (from {\it SPHERE} internal documentation) shows a model atmospheric spectrum for a 1 M$_{Jup}$ 10 Myr 
planet (solid black curve).  Simultaneous differential imagers can identify planets by comparing the flux in two adjacent 
narrow bands (like the H2 and H3 bands shown on the plot) and searching for the signature absorption drop.  To 
understand the sensitivity of this technique to planetary masses, one must first predict the expected planet flux drop 
between the two narrow-band filters.  While theoretical models predict such flux changes, transiting planet observations 
provide empirical data to test such assumptions. 
     Integral field spectrographs go one step further by imaging a potential star planet system as well as taking a low 
resolution (like R$\sim$50) spectrum at each point in the image.  A planet signal may be identified by running a correlation 
analysis between the observed spectrum and an expected planet spectrum.  An error in the prediction of expected planet 
spectra would lead to a degradation in sensitivity.  Transiting planet atmospheric studies would therefore be helpful in 
improving these correlation analyses. 

\begin{figure}[b]
\begin{center}
\includegraphics[width=5.4in]{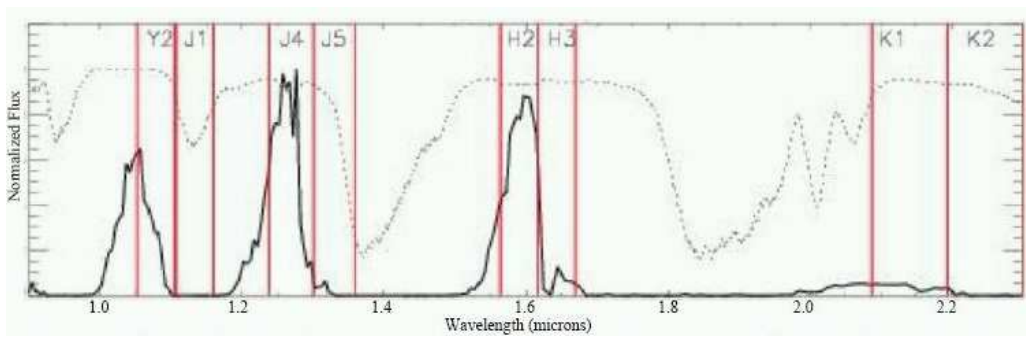}
\caption{A model 1M$_{Jup}$ planet spectrum (solid curve) with narrow band filters (vertical red lines) shown on top.  Figure from {\it SPHERE} internal documentation.} 
\label{fig1}
\end{center}
\end{figure}

     For direct-imaging detections, the planet mass 
must typically be derived indirectly via 
theoretical evolutionary models like the one shown in Figure 3.
Transiting 
planet data, in their capacity to help improve 
these evolutionary tracks, may therefore lead to 
better direct-imaging planet mass estimates.

\begin{figure}[b]
\begin{center}
\includegraphics[width=6in]{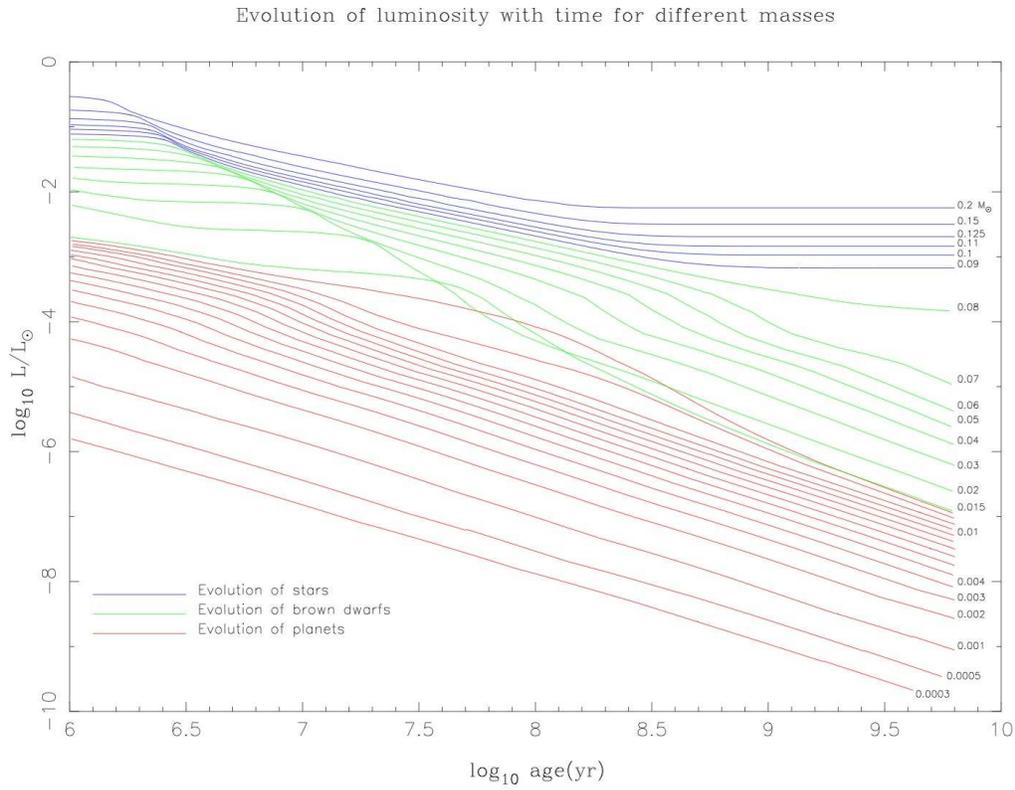}
\caption{Evolutionary cooling models by A. Burrows for M dwarfs (blue curves), brown dwarfs (green curves), and exo-solar giant planets (red curves).  Plot from http://www.astro.princeton.edu/$\sim$burrows.
}
\label{fig1}
\end{center}
\end{figure}

\section{Complementarity Between Transit and Direct-Imaging Planet Observations}
     While transiting planets are unlikely to be successfully observed with direct-imaging techniques, 
transiting planet systems nevertheless make appealing targets to search for wider orbit planets.  Indeed, the 
discovery (via direct-imaging) of an outer planet around a transiting planet system would make an exciting 
case to observe planet multiplicity through a large semi-major axis range.  As planet detections via these 
techniques reach a statistically significant number, the results should allow us a more complete picture of 
inward/outward migration, planet multiplicity, dynamical equilibrium, and general planet evolution.

\end{document}